\documentclass[aps,prd,twocolumn,superscriptaddress,showkeys]{revtex4}
\usepackage{amssymb,amsmath,graphicx,enumerate}
\usepackage{subfigure}
\usepackage{dcolumn,booktabs}
\usepackage{longtable}
\usepackage[dvipdfm,colorlinks=true,citecolor=blue,pdfstartview=FitH]{hyperref}

\def\bea{\begin{equation}}
\def\eea{\end{equation}}
\newcommand{\rt}{Regge trajectory}
\newcommand{\rts}{Regge trajectories}
\newcommand{\bfr}{{\bf r}}
\newcommand{\bfp}{{\bf p}}
\newcommand{\bfpa}{{|\bf p|}}
\newcommand{\gev}{{\rm GeV}}

\newcommand{\cltb}{$\bar{3}_c$}
\newcommand{\cltba}{\bar{3}_c}
\newcommand{\cls}{$6_c$}
\newcommand{\dqs}{$(qq')$}

\begin{document}
\title{Regge trajectories for the doubly heavy diquarks}
\author{Xia Feng}
\email{sxsdwxfx@163.com}
\affiliation{School of Physics and Information Engineering, Shanxi Normal University, Taiyuan 030031, China}
\author{Jiao-Kai Chen}
\email{chenjk@sxnu.edu.cn, chenjkphy@outlook.com}
\affiliation{School of Physics and Information Engineering, Shanxi Normal University, Taiyuan 030031, China}
\author{Jia-Qi Xie}
\email{1462718751@qq.com}
\affiliation{School of Physics and Information Engineering, Shanxi Normal University, Taiyuan 030031, China}
\date{\today}

\begin{abstract}
The concept of diquark is important for understanding hadron structure and high-energy particle reactions.
We attempt to apply the Regge trajectory approach to the doubly heavy diquarks. We present a method for determining the parameters in the diquark Regge trajectory. The spectra of diquarks $(cc)$, $(bb)$, and $(bc)$ are obtained by using the {\rt} approach and are found to agree with other theoretical results.
The diquark Regge trajectory becomes a new and very simple approach for estimating the spectra of diquarks.
\end{abstract}

\maketitle


\section{Introduction}
The concept of diquark is important for understanding hadron structure and high-energy particle reactions \cite{Gell-Mann:1964ewy,Lichtenberg:1967zz,
Anselmino:1992vg,Jaffe:2004ph,Barabanov:2020jvn,Selem:2006nd,
Wilczek:2004im,Jaffe:2003sg}.
In the diquark picture, baryons are bound states of a diquark and one quark \cite{Lichtenberg:1967zz,Ida:1966ev}. A tetraquark is composed of a diquark and an antidiquark \cite{Liu:2019zoy,Jaffe:1976ig,Esposito:2016noz}. Pentaquarks are clusters of two diquarks and one antiquark \cite{Liu:2019zoy,Jaffe:2004ph,Esposito:2016noz} or of one diquark and one triquark \cite{Lebed:2015tna}.
Phenomenology suggests that diquark correlations might play a material role in the formation of exotic tetraquarks and pentaquarks \cite{Jaffe:2004ph,Liu:2019zoy,Esposito:2016noz,Olsen:2017bmm,Guo:2019twa}.
Diquark substructure affects the static properties of baryons, tetraquarks and pentaquarks.

Various authors have studied the spectra of diquarks. In Ref. \cite{Gershtein:2000nx}, the spectra of the doubly heavy diquarks $(QQ')$ are calculated by using the Schr\"{o}dinger equation. In Ref. \cite{Ebert:2002ig}, the mass spectra of the doubly heavy diquarks are obtained by the quasipotential equation of the Schr\"{o}dinger type. In Ref. \cite{Gutierrez-Guerrero:2021fuj}, the masses of different kinds of diquarks are calculated within a nonrelativistic potential model. In Refs. \cite{Gutierrez-Guerrero:2021rsx,Yin:2021uom,
Li:2019ekr,Yu:2006ty,Wang:2005tq,Cahill:1987qr,Maris:2002yu}, the spectra of diquarks are computed by using the Bethe-Salpeter equation. In Refs. \cite{Lu:2017meb,Ferretti:2019zyh}, the diquark masses are calculated by the spinless Salpeter equation. In Ref. \cite{Hess:1998sd}, the diquark masses are obtained from lattice QCD.

The {\rt} is one of the effective approaches for studying hadron spectra \cite{Regge:1959mz,Chew:1962eu,Nambu:1978bd,Ademollo:1969nx,Baker:2002km,
Brodsky:2006uq,Forkel:2007cm,Filipponi:1997vf,Anisovich:2000kxa,brau:04bs,
Brisudova:1999ut,Masjuan:2012gc,Chen:2014nyo,Guo:2008he,Ebert:2009ub,
Feng:2023ynf,Lovelace:1969se,Irving:1977ea,Collins:1971ff,Inopin:1999nf,Afonin:2014nya,Badalian:2019dny,
MartinContreras:2020cyg,Sonnenschein:2018fph,Chen:2023web,Chen:2023djq,Chen:2022flh}. 
In the present work, we attempt to apply this approach to investigate the diquark spectra, even though diquarks are colored states and not physical \cite{Jaffe:2004ph}. We use the term diquark {\rt} following the same nomenclature as the hadron {\rt} \cite{note}. Our focus in this work is on the doubly heavy diquarks $(cc)$, $(bb)$ and $(bc)$.

The paper is organized as follows: In Sec. \ref{sec:rgr}, the {\rt} relation is obtained from the quadratic form of the spinless Salpeter-type equation (QSSE). In Sec. \ref{sec:rtdiquark}, we investigate the {\rts} for the doubly heavy diquarks. The conclusions are presented in Sec. \ref{sec:conc}.

\section{{\rt} relations}\label{sec:rgr}

We apply the ansatz \cite{Chen:2022flh} to describe the diquark spectra \bea\label{massf}
M=\beta_x(x+c_{0x})^{\nu}+m_R,\,(x=l,\,n_r)
\eea
where $M$ is the diquark mass, $l$ is the orbital angular momentum and $n_r$ is the radial quantum number. $\beta_x$, $\nu$ and $m_R$ can be determined by employing the QSSE, see Eq. (\ref{massform}). $c_0$ varies with different trajectories.

\subsection{QSSE}

The QSSE reads as \cite{Baldicchi:2007ic,Baldicchi:2007zn,Brambilla:1995bm,
chenvp,chenrm,Chen:2018hnx,Chen:2018bbr,Chen:2018nnr,Chen:2021kfw}
\begin{eqnarray}\label{qsse}
M^2\Psi({\bfr})=M_0^2\Psi({\bfr})+{\mathcal U}\Psi({\bfr}),\quad M_0=\omega_1+\omega_2,
\end{eqnarray}
where $\Psi$ is the diquark wave function, $\omega_1$ and $\omega_2$ are the square-root operators of the relativistic kinetic energy of quark $q$ and quark $q'$, respectively,
\bea\label{omega}
\omega_i=\sqrt{m_i^2+{\bf p}^2}=\sqrt{m_i^2-\Delta},
\eea
\bea\label{potu}
{\mathcal U}=M_0V_{qq'}+V_{qq'}M_0+V_{qq'}^2.
\eea
$m_1$ and $m_2$ are the effective masses of quark $q$ and quark $q'$, respectively.
In $SU_c(3)$, there is attraction between quark pairs $(qq')$ in the color antitriplet channel, and this is just twice weaker than in the color singlet $q\bar{q}'$ in the one-gluon exchange approximation \cite{Esposito:2016noz}.
Only the {\cltb} representation of $SU_c(3)$ is considered in the present work and the {\cls} representation \cite{Weng:2021hje,Praszalowicz:2022sqx} is not considered. In the case of diquarks in a color {\cltb} state, the relation \cite{Ebert:2002ig,Gershtein:2000nx}
\bea
V_{qq'}=\frac{V_{q\bar{q}'}}{2}
\eea
is used for the quark-quark interaction.
The quark-antiquark interaction in the mesons adopts the Cornell potential \cite{Eichten:1974af},
\bea\label{potv}
V_{q\bar{q}'}(r)=-\frac{\alpha}{r}+{\sigma}r+C,
\eea
where the first term is the color Coulomb potential parameterized by the coupling strength $\alpha$. The second term is the linear confining potential, and $\sigma$ is the string tension. $C$ is a fundamental parameter \cite{Gromes:1981cb,Lucha:1991vn}.

\subsection{{\rts} obtained from the QSSE}
For the heavy-heavy diquarks, $m_{1},m_2{\gg}{\bfpa}$, Eq. (\ref{qsse}) reduces to
\begin{eqnarray}\label{qssenr}
M^2\Psi({\bfr})&=&\left[(m_1+m_2)^2+\frac{m_1+m_2}{\mu}{\bfp}^2\right]\Psi({\bfr})\nonumber\\
&&+2(m_1+m_2)V\Psi({\bfr}),
\end{eqnarray}
where
\bea
\mu=m_1m_2/(m_1+m_2).
\eea
By employing the Bohr-Sommerfeld quantization approach \cite{Brau:2000st,brsom} and using Eqs. (\ref{potv}) and (\ref{qssenr}), we can obtain the orbital and radial {\rts} \cite{Chen:2018hnx,Chen:2021kfw},
\begin{align}\label{rtnrs1}
M^2{\sim}&3(m_1+m_2)\left(\frac{\sigma'^2}{\mu}\right)^{1/3}l^{2/3} \quad (l{\gg}n_r),\nonumber\\
M^2{\sim}&(3\pi)^{2/3}(m_1+m_2)\left(\frac{\sigma'^2}{\mu}\right)^{1/3}{n_r}^{2/3}\quad (n_r{\gg}l),
\end{align}
where $\sigma'=\sigma/2$.
By considering the constant term of $m_1$ and $m_2$, as well as the omitted term proportional to $n_r^{4/3}$, we can obtain (\ref{massf}) from Eq. (\ref{rtnrs1}) with the following parameters \cite{Chen:2022flh}
\bea\label{massform}
\nu=2/3,\; \beta_x=c_{fx}c_xc_c,\;m_R=m_1+m_2+\epsilon_c.
\eea
Here, $\epsilon_c$ is a constant which is a part of the interaction energy. Usually, $\epsilon_c=C/2$ where $C$ is given by Eq. (\ref{potv}). The constants $c_{x}$ and $c_c$ are
\bea\label{cxcons}
c_c=\left(\frac{\sigma'^2}{\mu}\right)^{1/3},\quad c_l=\frac{3}{2},\quad c_{n_r}=\frac{\left(3\pi\right)^{2/3}}{2}.
\eea
Both $c_{fl}$ and $c_{fn_r}$ are theoretically equal to 1 and are fitted in practice.
In Eqs. (\ref{massform}) and (\ref{cxcons}), $m_1$, $m_2$, $\epsilon_c$, $c_x$, $c_{fx}$ and $\sigma'$ are universal for the heavy-heavy diquarks. $c_{0x}$ is determined by fitting a given {\rt}.
The simple formula (\ref{massf}) with (\ref{massform}) and (\ref{cxcons}) can give results which are consistent with other theoretical predictions; see Sec. \ref{sec:rtdiquark}.

If the confining potential is $V_{qq'}(r)={\sigma'}r^{a}$ ($a > 0$), Eq. (\ref{massf}) becomes
\bea
M=m_R+c_{fx}c_xc_c(x+c_{0x})^{2a/(a+2)}.
\eea
$c_x$ and $c_c$ are
\begin{align}\label{eq11}
c_l&=\left(1+\frac{a}{2}\right)
\left(\frac{1}{a}\right)^{a/(a+2)},\nonumber\\
c_{n_r}&=\left(\frac{1}{2}\right)^{a/(a+2)}
\left(\frac{a\pi}{B(1/a,3/2)}\right)^{2a/(a+2)},\nonumber\\
c_c&=\left(\frac{\sigma'^2}{\mu^{a}}\right)^{1/(a+2)},
\end{align}
where $B(x,y)$ denotes the beta function \cite{Gradshteyn:book1980}.
Different forms of kinematic terms corresponding to different energy regions will yield different behaviors of the {\rts} \cite{Chen:2021kfw,Chen:2022flh}. ${\bf p}$ and $r^a$ lead to $M{\sim}x^{a/(a+1)}$ while ${\bf p}^2$ and $r^a$ result in $M{\sim}x^{2a/(a+2)}$ $(x=l,\,n_r)$. For the heavy-heavy systems, the {\rts} plotted in the $\left(M,\,x\right)$ plane will exhibit an upward convexity if $a>2$, a linear relationship  if $a=2$, and a downward concavity if $a<2$.

Equation (\ref{massf}) with (\ref{massform}) can lead to another form of the {\rt} \cite{Chen:2022flh,Chen:2023djq,note}
\bea\label{reglike}
(M-m_R)^2=\alpha_x(x+c_{0x})^{4/3}\quad (x=l,\,n_r),
\eea
where $\alpha_x=c^2_{fx}c^2_{x}c_c^2$. Equation (\ref{reglike}) is similar to the formulas in Refs. \cite{Badalian:2019dny,Afonin:2014nya}.
For the confining potential $V_{qq'}(r)={\sigma'}r^{a}$ ($a > 0$), Eq. (\ref{reglike}) becomes
\bea
(M-m_R)^2=\alpha_x(x+c_{0x})^{4a/(a+2)}.
\eea
For the heavy-heavy systems, the {\rts} plotted in the $\left((M-m_R)^2,\,x\right)$ plane will be convex upwards if $a>2/3$, be linear if $a=2/3$, and be concave downwards if $a<2/3$ \cite{Chen:2018bbr}.

\section{{\rts} for the doubly heavy diquarks}\label{sec:rtdiquark}

In this section, the {\rts} for the doubly heavy diquarks $(cc)$, $(bb)$ and $(bc)$ are investigated by using Eq. (\ref{massf}).

\subsection{Preliminary}

The state of diquark $(qq')$ is denoted as $[qq']^{{\cltba}}_{n^{2s+1}l_j}$ or $\{qq'\}^{{\cltba}}_{n^{2s+1}l_j}$, where $\{qq'\}$ and $[qq']$ indicate the permutation symmetric and antisymmetric flavor wave functions, respectively. $n=n_r+1$, $n_r=0,1,\cdots$, where $n_r$ is the radial quantum number. $s$ is the total spin of two quarks, $l$ is the orbital quantum number, and $j$ is the spin of the diquark $(qq')$. {\cltb} denotes the color antitriplet state of diquark.

For the diquarks $(cc)$ and $(bb)$, the states with antisymmetric flavor wave function do not exist. For the diquark $(bc)$, both $[bc]$ and $\{bc\}$ exist; see the Appendix and Table \ref{tab:dqstates} for more discussions.

Using the formula (\ref{massf}) with (\ref{massform}) and (\ref{cxcons}) [$\sigma'=\sigma$ and $\epsilon_c=C$ for mesons] and the PDG averaged masses from the Particle Data Group \cite{ParticleDataGroup:2022pth}, we fit the radial and orbital {\rts} for the charmonia, bottomonia and the bottom-charmed mesons, respectively. The quality of a fit is measured by the quantity $\chi^2$ defined by \cite{Sonnenschein:2014jwa}
\bea
\chi^2=\frac{1}{N-1}\sum^{N}_{i=1}\left(\frac{M_{fi}-M_{ei}}{M_{ei}}\right)^2,
\eea
where $N$ is the number of points on the trajectory, $M_{fi}$ is fitted value and $M_{ei}$ is the experimental value or the theoretical value of the $i$-th particle. The parameters are determined by minimizing $\chi^2$.
We use the following parameter values \cite{Ebert:2002ig,Faustov:2021qqf} to fit the {\rts} for the doubly heavy diquarks,
\begin{align}\label{param}
m_b&=4.88\; {\gev},\quad m_c=1.55\; {\gev},\nonumber\\
\sigma&=0.18\; {\gev^2},\quad C=-0.30\; {\gev},\nonumber\\
c_{fl}&=1.17,\quad c_{fn_r}=1.
\end{align}
The values of $m_b$, $m_c$, $\sigma$ and $C$ are taken directly from \cite{Ebert:2002ig,Faustov:2021qqf}. $c_{fl}$ and $c_{fn_r}$ are obtained by fitting the {\rts} for the doubly heavy mesons.
These parameters are universal for all doubly heavy diquark {\rts}. There is only one free parameter $c_{0x}$ in (\ref{massf}) or (\ref{reglike}), which is determined by fitting the corresponding meson {\rts} and varies with different diquark {\rts}. There is not compelling reason why $c_{0x}$ obtained by fitting the meson {\rts} can be used directly to calculate the diquark {\rts}. We use this method as a provisional method to determine $c_{0x}$ before finding a better one. It validates this method that the fitted results for the diquarks $(cc)$, $(bb)$, and $(bc)$ agree with the theoretical values obtained by using other approaches, see the discussions in the following subsections.

\begin{figure}[!phtb]
\centering
\subfigure[]{\label{figcca}\includegraphics[scale=0.75]{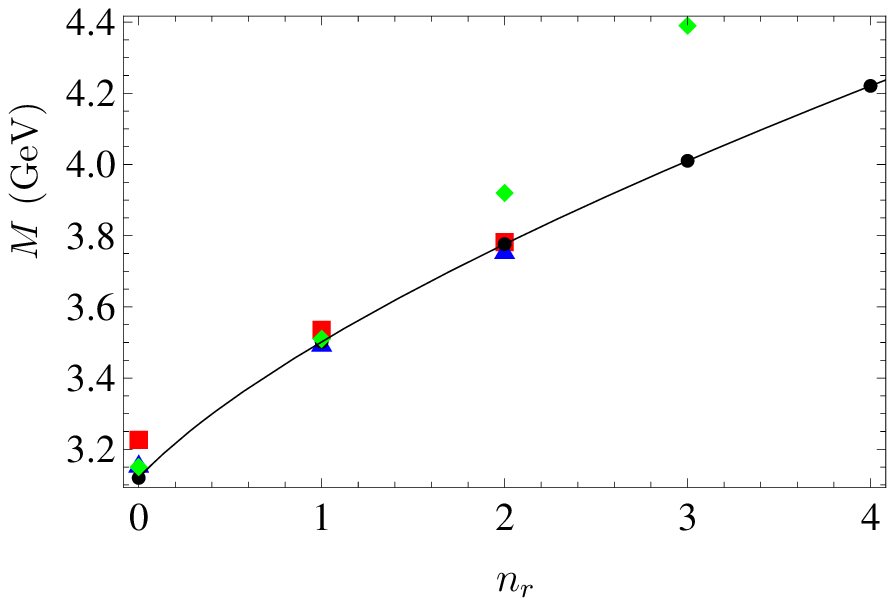}}
\subfigure[]{\label{figccb}\includegraphics[scale=0.75]{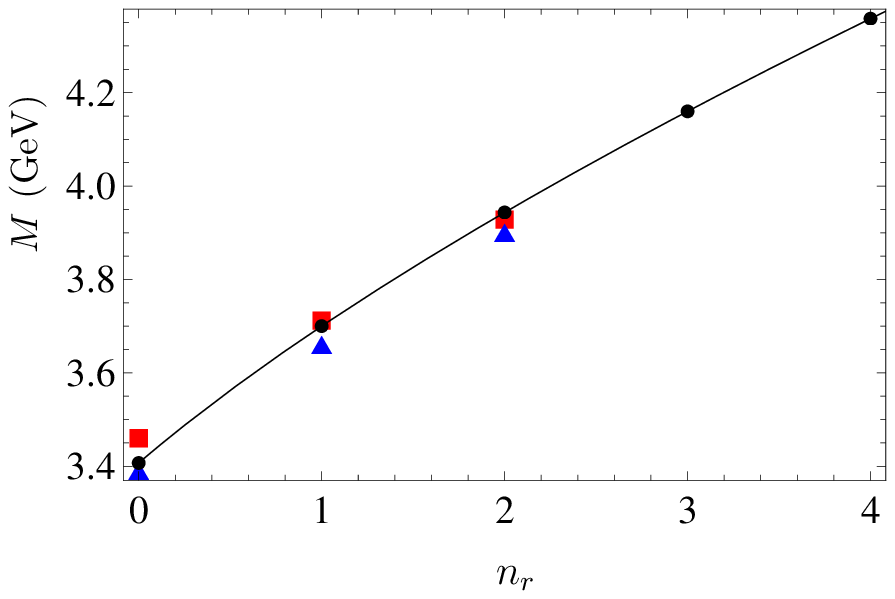}}
\subfigure[]{\label{figccc}\includegraphics[scale=0.75]{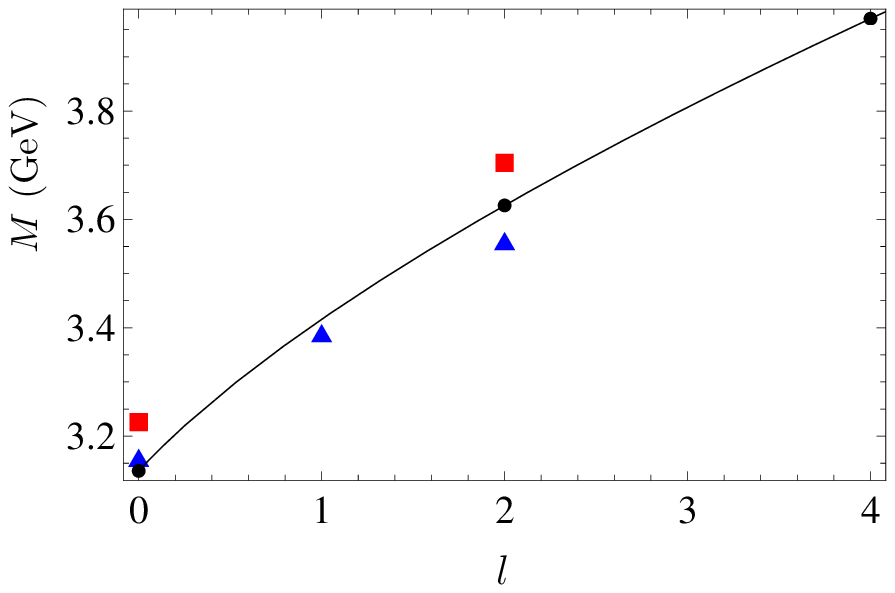}}
\caption{The radial and orbital {\rts} for the $(cc)$ diquark. (a) Radial {\rt} for the diquark $\{cc\}^{\bar{3}_c}_{1^3s_1}$ state. (b) Radial {\rt} for the diquark $\{cc\}^{\bar{3}_c}_{1^1p_1}$ state. (c) Orbital {\rt} for the $\{cc\}^{\bar{3}_c}_{1^3s_1}$ state. The data for the red filled squares are from EFMG or FG. The data for the blue filled triangles are from GKLO. The data for the green diamonds are from GAR. The black lines and the black dots are our results. The data are listed in Tables \ref{tab:ccr} and \ref{tab:ccl}.}\label{fig:dqcc}
\end{figure}

\subsection{{\rts} for the $(cc)$ diquark}

By using Eq. (\ref{massf}) with (\ref{massform}), (\ref{cxcons}) and (\ref{param}) [$\sigma'=\sigma$ and $\epsilon_c=C$ for mesons] to fit the radial $J/\psi(1S)$ and $h_c(1P)$ {\rts}, we have $c_{0n_r}=0.205$ and $c_{0n_r}=0.907$, respectively.
Fitting the orbital $\eta_c(1S)$ and $J/\psi(1S)$ {\rts} gives $c_{0l}=0.188$ and $c_{0l}=0.337$, respectively.
To calculated the masses of the $\{cc\}^{{\cltba}}_{1^{3}d_1}$ and $\{cc\}^{{\cltba}}_{1^{3}d_2}$ states, we fit the orbital $\chi_{c0}(1P)$ and $\chi_{c1}(1P)$ {\rts} and obtain $c_{0l}=1.016$, $c_{0l}=1.231$, respectively.

\begin{table}[!phtb]
\caption{Comparison of the theoretical values (in {\gev}) for the diquark $(cc)$. The data from Ref. \cite{Gershtein:2000nx} are without spin-dependent splittings. $n=n_r+1$, $n_r=0,1,\cdots$. $n_r$ is the radial quantum number. $s$ is the total spin of two quarks, $l$ is the orbital quantum number and $j$ is the spin of diquark. The acronyms are from the
initial letters of the last name of authors.}  \label{tab:ccr}
\centering
\begin{tabular*}{0.47\textwidth}{@{\extracolsep{\fill}}ccccc@{}}
\hline\hline
State ($n^{2s+1}l_j$)   &  EFGM \cite{Ebert:2002ig}   &  GKLO \cite{Gershtein:2000nx}  & GAR \cite{Gutierrez-Guerrero:2021fuj}  & Our results   \\
\hline
$1^3s_1$        & 3.226       &3.16         & 3.15 & 3.12 \\
$2^3s_1$        & 3.535       &3.50         & 3.51 & 3.50 \\
$3^3s_1$        & 3.782       &3.76         & 3.92 & 3.78 \\
$4^3s_1$        &             &             & 4.39 & 4.01 \\
$5^3s_1$        &             &             &      & 4.22  \\
$1^1p_1$    &3.460     &3.39                &  & 3.41 \\
$2^1p_1$    &3.712     &3.66                &  & 3.70\\
$3^1p_1$    &3.928     &3.90                &  & 3.94\\
$4^1p_1$    &          &                    &  & 4.16 \\
$5^1p_1$    &          &                    &  & 4.36 \\
\hline\hline
\end{tabular*}
\end{table}

\begin{table}[!phtb]
\caption{The theoretical values (in {\gev}) for the $(cc)$ diquark. The data from Ref. \cite{Gershtein:2000nx} are without spin-dependent splittings. $\times$ denotes the nonexistent states. The acronyms are from the
initial letters of the last name of authors.}  \label{tab:ccl}
\centering
\begin{tabular*}{0.47\textwidth}{@{\extracolsep{\fill}}cccc@{}}
\hline\hline
State ($n^{2s+1}l_j$)   &  FG \cite{Faustov:2021qqf}   & GKLO \cite{Gershtein:2000nx}   & Our results \\
\hline
$1^3s_1$        &3.226     &3.16     & 3.14    \\
$1^3p_2$        &          &3.39     & 3.42($\times$)  \\
$1^3d_3$        &3.704     &3.56     & 3.63  \\
$1^3f_4$        &          &         & 3.81($\times$)    \\
$1^3g_5$        &          &         & 3.97    \\
\hline\hline
\end{tabular*}
\end{table}

Substituting the values in Eq. (\ref{param}) and the obtained $c_{0x}$ into  (\ref{massform}) and (\ref{cxcons}), Eq. (\ref{massf}) is determined [$\sigma'=\sigma/2$ and $\epsilon_c=C/2$ for diquarks]. We use  (\ref{massf}) to calculate the masses of diquark $(cc)$, which are listed in Tables \ref{tab:ccr}, \ref{tab:ccl} and \ref{tab:cpmass}.
As we fit the parameter $c_{0x}$, the spin-dependent effects are considered in fact. For example, we fit the orbital $J/\psi(1S)$ {\rt}, then the obtained ${c_{0l}}$ is used for the orbital $\{cc\}^{{\cltba}}_{1^{3}s_1}$ {\rt}.
Here, we do not consider the $^3(j-1)_{j}-^3(j+1)_j$ mixing of diquarks which is similar to the mixing of different wave states of charmonium and bottomonium \cite{Eichten:2007qx}.
The mass of state $\{cc\}^{{\cltba}}_{2^{3}s_1}$ is estimated by using the radial $\{cc\}^{{\cltba}}_{1^{3}s_1}$ {\rt}, $m(\{cc\}^{{\cltba}}_{2^{3}s_1})=3.535$ {\gev}. The mass of state $\{cc\}^{{\cltba}}_{1^{3}d_1}$ is computed by using the orbital $\{cc\}^{{\cltba}}_{1^{3}p_0}$ {\rt}, $m(\{cc\}^{{\cltba}}_{1^{3}d_1})=3.56$ {\gev}.

The spectrum from GKLO \cite{Gershtein:2000nx} is without spin-dependent splittings. The symbol $\times$ in Table \ref{tab:ccl} denotes the nonexisting states according to the Pauli exclusion principle; see the Appendix for more details.

Our results obtained by the {\rt} approach are consistent with other theoretical results, as shown in Tables \ref{tab:ccr}, \ref{tab:ccl}, and \ref{tab:cpmass} and Fig. \ref{fig:dqcc}. As depicted in Fig. \ref{fig:dqcc}, our results are in agreement with EFGM \cite{Ebert:2002ig} and GKLO\cite{Gershtein:2000nx}. The behavior of {\rt} is different from the data from GAR \cite{Gutierrez-Guerrero:2021fuj}; see Fig. \ref{figcca}.

\subsection{{\rts} for the $(bb)$ diquark}

\begin{figure}[!phtb]
\centering
\subfigure[]{\label{figbba}\includegraphics[scale=0.75]{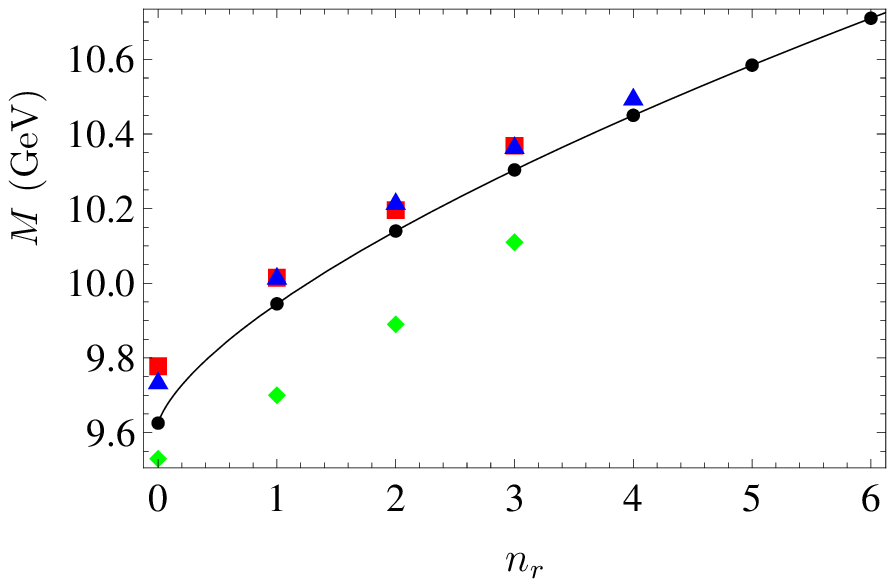}}
\subfigure[]{\label{figbba}\includegraphics[scale=0.75]{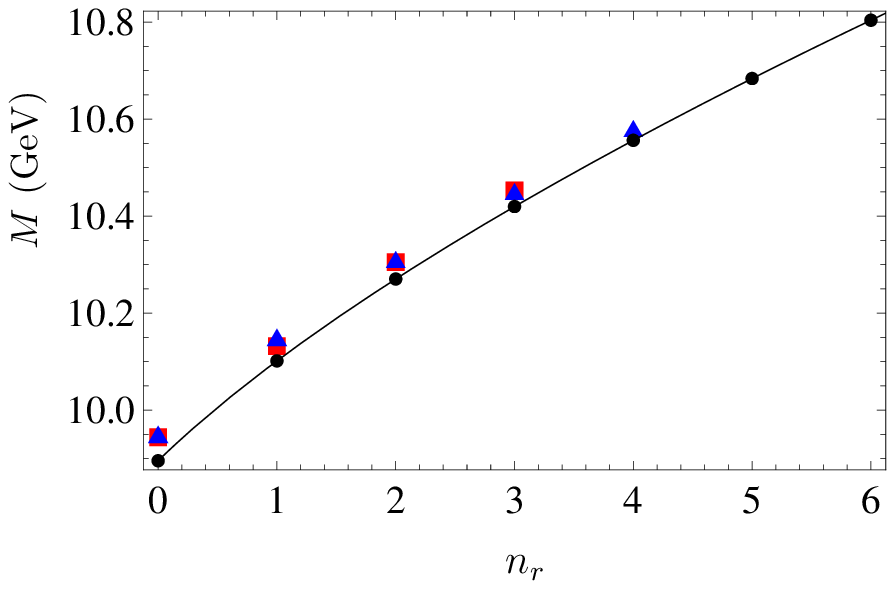}}
\subfigure[]{\label{figbba}\includegraphics[scale=0.75]{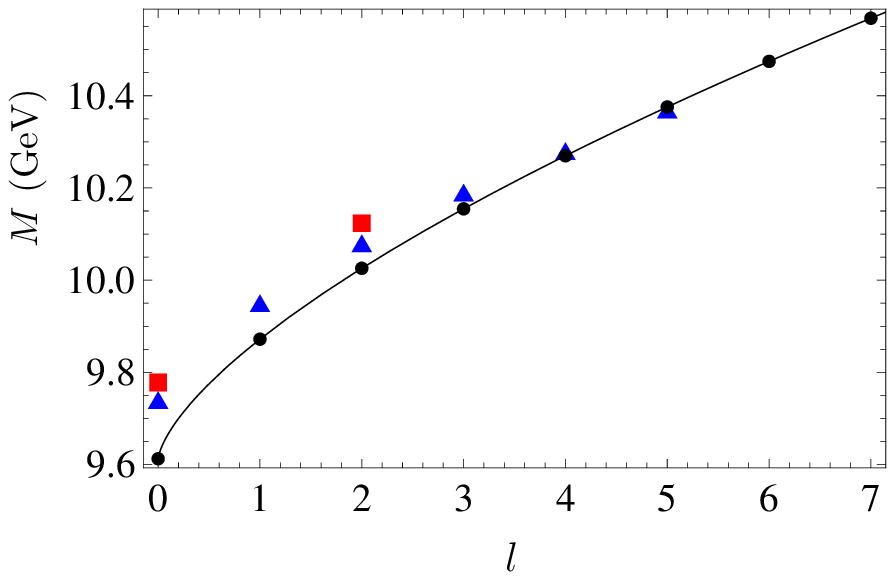}}
\caption{The radial and orbital {\rts} for the $(bb)$ diquark. (a) Radial {\rt} for the $\{bb\}^{\bar{3}_c}_{1^3s_1}$ state. (b) Radial {\rt} for the $\{bb\}^{\bar{3}_c}_{1^1p_1}$ state. (c) Orbital {\rt} for the $\{bb\}^{\bar{3}_c}_{1^3s_1}$ state. The data for the red filled squares are from EFGM or FG. The data for the blue filled triangles are from GKLO. The data for the green diamonds are from GAR. The black lines and the black dots are our results. The data are listed in Tables \ref{tab:bbr} and \ref{tab:bbl}.}\label{fig:dqbb}
\end{figure}

By using Eq. (\ref{massf}) in conjunction with (\ref{massform}), (\ref{cxcons}) and (\ref{param}) to fit the radial $\Upsilon(1S)$ and $h_b(1P)$ {\rts}, we have $c_{0n_r}=0.01$ and $c_{0n_r}=0.795$, respectively.
Fitting the orbital $\eta_b(1S)$ and $\Upsilon(1S)$ {\rts} gives $c_{0l}=0$, $c_{0l}=0.001$, respectively.
To calculate the masses of the $\{bb\}^{{\cltba}}_{1^{3}d_1}$ and $\{bb\}^{{\cltba}}_{1^{3}d_2}$ states, we fit the orbital $\chi_{b0}(1P)$ {\rt} and the radial $\chi_{b1}(1P)$ {\rt} and obtain $c_{0l}=0.942$, $c_{0n_r}=0.786$, respectively.

Using Eq. (\ref{massf}) with (\ref{massform}), (\ref{cxcons}), (\ref{param}) and the obtained $c_{0x}$, we calculate the masses of diquark $(bb)$. The masses are listed in Tables \ref{tab:bbr}, \ref{tab:bbl} and \ref{tab:cpmass}. The corresponding {\rts} are in Fig. \ref{fig:dqbb}. As shown in Tables \ref{tab:bbr}, \ref{tab:bbl}, and \ref{tab:cpmass} and Fig. \ref{fig:dqbb}, our results are in accordance with other theoretical results.

Similar to the $(cc)$ diquark case, we do not consider the $^3(j-1)_{j}-^3(j+1)_j$ mixing of $(bb)$ diquarks.
The mass of state $\{bb\}^{{\cltba}}_{2^{3}s_1}$ is estimated by using the radial $\{bb\}^{{\cltba}}_{1^{3}s_1}$ {\rt}, $m(\{bb\}^{{\cltba}}_{2^{3}s_1})=10.015$ {\gev}. The mass of state $\{bb\}^{{\cltba}}_{1^{3}d_1}$ is computed by using the orbital $\{bb\}^{{\cltba}}_{1^{3}p_0}$ {\rt}, $m(\{bb\}^{{\cltba}}_{1^{3}d_1})=10.02$ {\gev}.

\begin{table}[!phtb]
\caption{Same as Table \ref{tab:ccr} except for the diquark $(bb)$.}  \label{tab:bbr}
\centering
\begin{tabular*}{0.47\textwidth}{@{\extracolsep{\fill}}ccccc@{}}
\hline\hline
State ($n^{2s+1}l_j$)  &  EFGM \cite{Ebert:2002ig}   &  GAR \cite{Gutierrez-Guerrero:2021fuj}&  GKLO \cite{Gershtein:2000nx}   & Our results     \\
\hline
$1^3s_1$        & 9.778   &  9.53  & 9.74   & 9.63    \\
$2^3s_1$        & 10.015  &  9.70  & 10.02  & 9.95 \\
$3^3s_1$        & 10.196  &  9.89  & 10.22   & 10.14  \\
$4^3s_1$        & 10.369  &  10.11  & 10.37  & 10.30    \\
$5^3s_1$        &         &         & 10.50  & 10.45    \\
$6^3s_1$        &         &         &        & 10.58    \\
$1^1p_1$        & 9.944   &      &  9.95  & 9.90    \\
$2^1p_1$        & 10.132  &      &  10.15 & 10.10  \\
$3^1p_1$        & 10.305  &      &  10.31 & 10.27  \\
$4^1p_1$        & 10.453  &      &  10.45 & 10.42 \\
$5^1p_1$        &         &      &  10.58 & 10.56 \\
$6^1p_1$        &         &      &        & 10.68 \\
\hline\hline
\end{tabular*}
\end{table}

\begin{table}[!phtb]
\caption{Same as Table \ref{tab:ccl} except for the diquark $(bb)$.}  \label{tab:bbl}
\centering
\begin{tabular*}{0.47\textwidth}{@{\extracolsep{\fill}}ccccc@{}}
\hline\hline
State ($n^{2s+1}l_j$)  &  FG \cite{Faustov:2021qqf}    & GKLO \cite{Gershtein:2000nx}   & Our results  \\
\hline
$1^3s_1$        & 9.778      &  9.74  & 9.61     \\
$1^3p_2$        &            &  9.95  & 9.87($\times$)  \\
$1^3d_3$        & 10.123     &  10.08 & 10.03 \\
$1^3f_4$        &            &  10.19 & 10.15($\times$) \\
$1^3g_5$        &            &  10.28 & 10.27 \\
$1^3h_6$        &            &  10.37 & 10.38($\times$) \\
\hline\hline
\end{tabular*}
\end{table}

\subsection{{\rts} for the $(bc)$ diquark}

\begin{table}[!phtb]
\caption{Same as Table \ref{tab:ccr} except for the diquark $(bc)$.}  \label{tab:bcr}
\centering
\begin{tabular*}{0.47\textwidth}{@{\extracolsep{\fill}}cccc@{}}
\hline\hline
State ($n^{2s+1}l_j$) &  GAR \cite{Gutierrez-Guerrero:2021fuj}&  GKLO \cite{Gershtein:2000nx}   & Our results    \\
\hline
$1^3s_1$           &  6.38  & 6.48   & 6.42    \\
$2^3s_1$           &  6.66  & 6.79  & 6.75 \\
$3^3s_1$           &  6.97  & 7.01   & 6.99  \\
$4^3s_1$           &  7.30  &      & 7.20    \\
$5^3s_1$           &         &      & 7.38    \\
$1^1s_0$           &  6.33  & 6.48   & 6.38    \\
$2^1s_0$           &  6.60  & 6.79  & 6.73 \\
$3^1s_0$           &  6.92  & 7.01   & 6.98  \\
$4^1s_0$           &  7.25  &      & 7.18    \\
$5^1s_0$           &         &      & 7.37    \\
$1^3p_0$           &         & 6.69 & 6.64    \\
$2^3p_0$           &         & 6.93 & 6.90    \\
$3^3p_0$           &         & 7.13 & 7.12    \\
$4^3p_0$           &         &      & 7.31    \\
$5^3p_0$           &         &      & 7.48    \\
$1^3d_1$           &         & 6.85 & 6.84    \\
$2^3d_1$           &         & 7.05 & 7.07    \\
$3^3d_1$           &         &      & 7.26    \\
$4^3d_1$           &         &      & 7.44    \\
$5^3d_1$           &         &      & 7.61    \\
\hline
\hline
\end{tabular*}
\end{table}

\begin{table}[!phtb]
\caption{Same as Table \ref{tab:ccl} except for the diquark $(bc)$.}  \label{tab:bcl}
\centering
\begin{tabular*}{0.47\textwidth}{@{\extracolsep{\fill}}ccc@{}}
\hline\hline
State ($n^{2s+1}l_j$) & GKLO \cite{Gershtein:2000nx}   & Our results  \\
\hline
$1^1s_0$         &  6.48  & 6.38  \\
$1^1p_1$         &  6.69  & 6.65  \\
$1^1d_2$         &  6.85  & 6.84 \\
$1^1f_3$         &  6.97  & 7.00 \\
$1^1g_4$         &  7.09  & 7.14 \\
$1^1h_5$         &  7.19  & 7.28 \\
$1^3s_1$         &  6.48  & 6.41     \\
$1^3p_2$         &  6.69  & 6.67  \\
$1^3d_3$         &  6.85  & 6.85 \\
$1^3f_4$         &  6.97  & 7.01 \\
$1^3g_5$         &  7.09  & 7.16 \\
$1^3h_6$         &  7.19  & 7.29 \\
\hline\hline
\end{tabular*}
\end{table}

\begin{figure*}[!phtb]
\centering
\subfigure[]{\label{figbca}\includegraphics[scale=0.65]{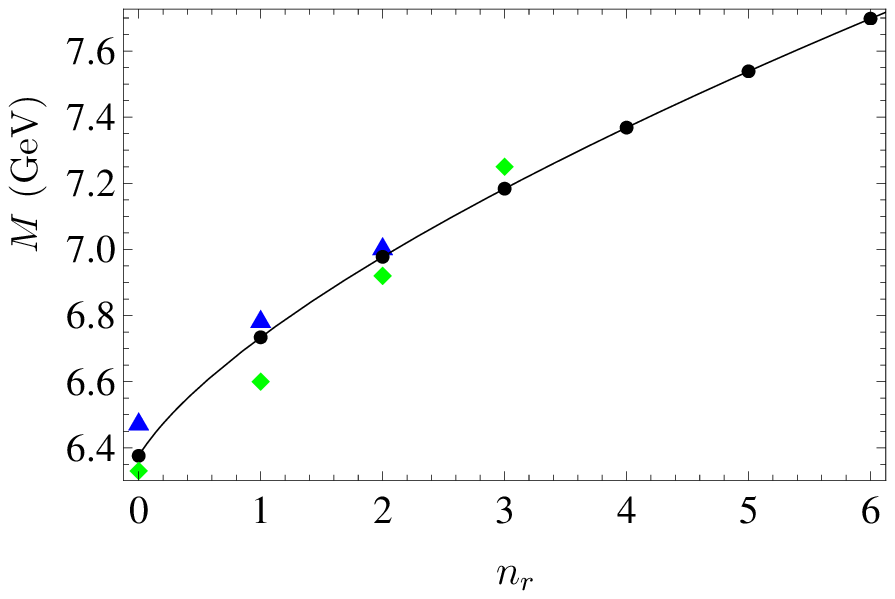}}
\subfigure[]{\label{figbca}\includegraphics[scale=0.65]{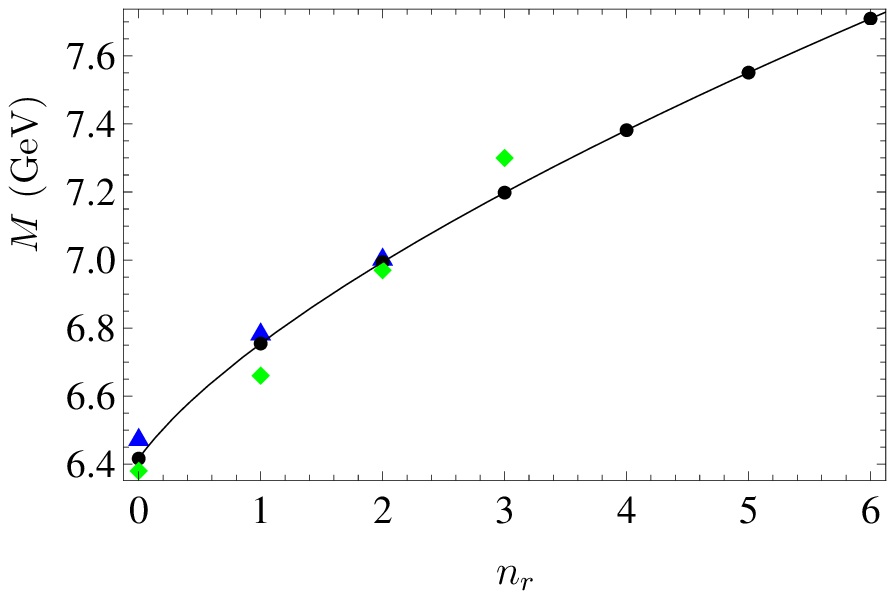}}
\subfigure[]{\label{figbca}\includegraphics[scale=0.65]{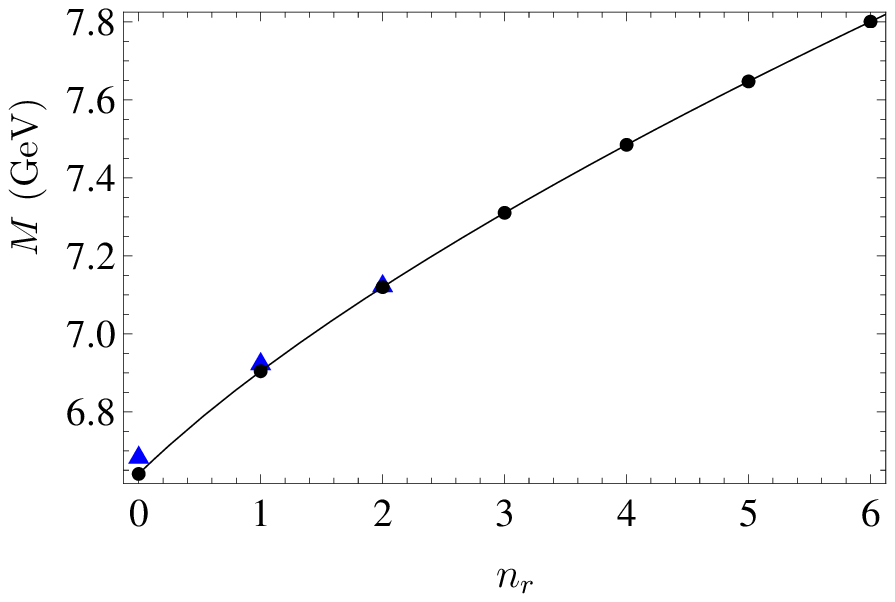}}
\subfigure[]{\label{figbca}\includegraphics[scale=0.65]{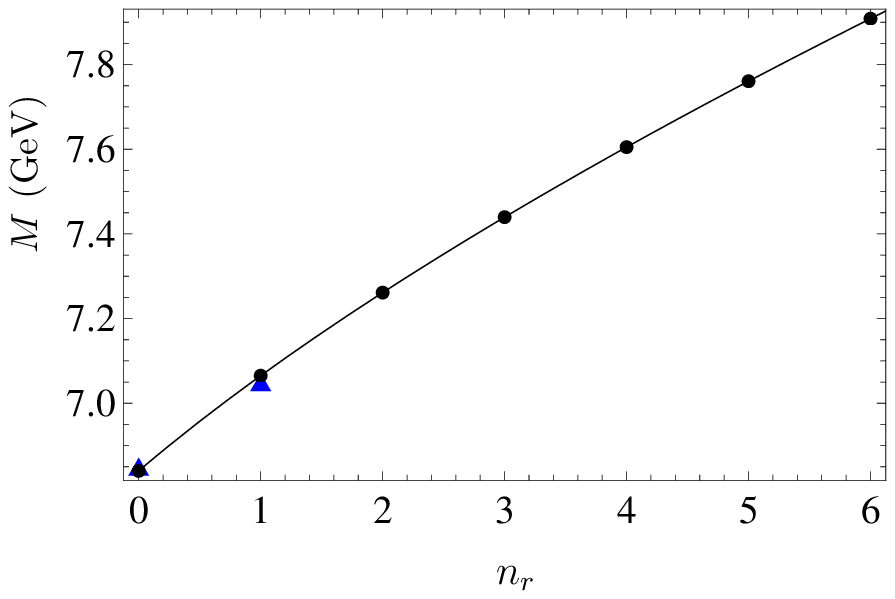}}
\subfigure[]{\label{figbca}\includegraphics[scale=0.65]{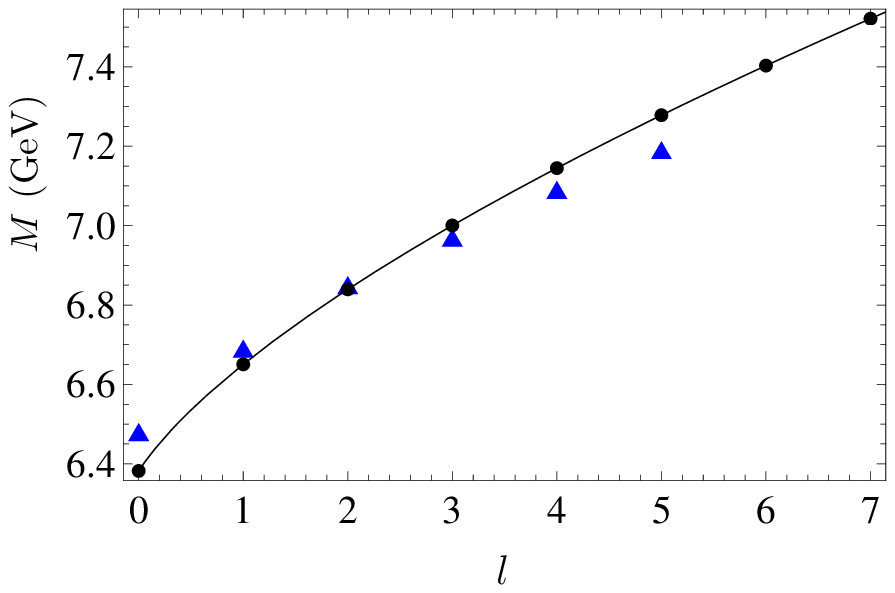}}
\subfigure[]{\label{figbca}\includegraphics[scale=0.65]{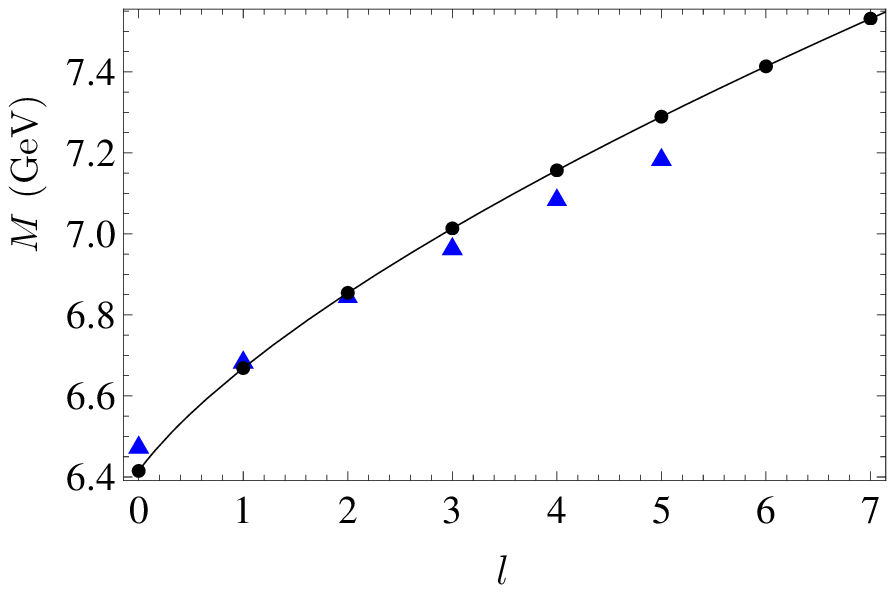}}
\caption{The radial and orbital {\rts} for the $(bc)$ diquark. (a), (b), (c) and (d) Radial {\rt} for the $\{bc\}^{\bar{3}_c}_{1^3s_0}$ state, the $\{bc\}^{\bar{3}_c}_{1^3s_1}$ state, the $\{bc\}^{\bar{3}_c}_{1^3p_0}$ and the $\{bc\}^{\bar{3}_c}_{1^3d_1}$ state, respectively. (e) and (f) Orbital {\rt} for the $\{bc\}^{\bar{3}_c}_{1^1s_0}$ state and the $\{bc\}^{\bar{3}_c}_{1^3s_1}$ state, respectively. The data for the blue filled triangles are from GKLO. The data for the green diamonds are from GAR. The black lines and the black dots are our results. The data are listed in Tables \ref{tab:bcr} and \ref{tab:bcl}.}\label{fig:dqbc}
\end{figure*}

\begin{table*}[!phtb]
\caption{Comparison of theoretical predictions for the masses of the doubly heavy diquarks (in {\gev}).}  \label{tab:cpmass}
\centering
\begin{tabular*}{0.8\textwidth}{@{\extracolsep{\fill}}cccccccc@{}}
\hline\hline
 $j^P$ & Diquark  &Our results&FG \cite{Faustov:2021qqf}, FGS \cite{Faustov:2020qfm}
 & GKLO \cite{Gershtein:2000nx} & GPB \cite{Gutierrez-Guerrero:2021rsx}
 &YCRS \cite{Yin:2021uom} & F \cite{Ferretti:2019zyh} \\
\hline
$0^+$   &  $[bc]^{{\cltba}}_{1^1s_0}$
                  & 6.38      & 6.519  & 6.48 &6.35 &6.48 &6.599 \\
$0^-$   & $[bc]^{{\cltba}}_{1^3p_0}$
                   & 6.64     &        & 6.69 &6.47 &6.62 &  \\
$1^+$   & $\{cc\}^{{\cltba}}_{1^3s_1}$
                   & 3.14     &3.226   & 3.16 &3.22 &3.30 &3.329 \\
        & $\{cc\}^{{\cltba}}_{1^3d_1}$
                   & 3.56     &3.704   & 3.56 &     &     &  \\
        & $\{bb\}^{{\cltba}}_{1^3s_1}$
                   & 9.63     &9.778   & 9.74 &9.44 &9.68 &9.845  \\
        & $\{bb\}^{{\cltba}}_{1^3d_1}$
                   & 10.02   & 10.123 & 10.08&     &     &  \\
        & $\{bc\}^{{\cltba}}_{1^3s_1}$
                   & 6.42     & 6.526  & 6.48 &6.35 &6.50 &6.611  \\
        & $\{bc\}^{{\cltba}}_{1^3d_1}$
                   & 6.84     &        & 6.85 &     &     & \\
$1^-$   & $\{cc\}^{{\cltba}}_{1^1p_1}$
                   & 3.41     & 3.460  & 3.39 &3.42 &     &   \\
        &  $\{bb\}^{{\cltba}}_{1^1p_1}$
                   & 9.90     & 9.944  & 9.95 &9.53 &     &  \\
        &  $\{bc\}^{{\cltba}}_{1^1p_1}$
                   & 6.65     &        & 6.69 &6.50 &     &   \\
\hline\hline
\end{tabular*}
\end{table*}

Although the excited states of the diquark $(bc)$ are unstable under the emission of soft gluons \cite{Gershtein:2000nx}, the authors provide the spectra of diquark $(bc)$. We obtain the orbital and radial diquark $(bc)$ {\rts} by fitting the corresponding $B_{c}$ {\rts}. Simultaneously, the spectra of diquark $(bc)$ are calculated.

There are scarce experimental data for the excited $B_{c}$ mesons. For the $B_c^+$ and $B_c(2S)^{\pm}$, the masses are from \cite{ParticleDataGroup:2022pth} while the theoretically predicted masses for other states are from Ref. \cite{Godfrey:2004ya}.

Because b quark and c quark are not identical, both $\{bc\}$ and $[bc]$ exist and there are more {\rts} for diquark $(bc)$ than that for diquarks $(cc)$ and $(bb)$.
By using Eq. (\ref{massf}) with (\ref{massform}), (\ref{cxcons}) and (\ref{param}) to fit the radial {\rts} for the $B_c(1^1S_0)$, $B_c(1^3S_1)$, $B_c(1^3P_0)$, $B_c(1^3D_1)$, we have $c_{0n_r}=0.107,\;0.182,\;0.783,\;1.516$, respectively.
Fitting the orbital {\rts} for the $B_c(1^1S_0)$ and $B_c(1^3S_1)$ gives $c_{0l}=0.169$ and $c_{0l}=0.257$, respectively.

Using Eq. (\ref{massf}) with (\ref{massform}), (\ref{cxcons}), (\ref{param}) and the obtained $c_{0x}$, we calculate the masses of diquark $(bc)$. The masses are listed in Tables \ref{tab:bcr}, \ref{tab:bcl} and \ref{tab:cpmass}. The corresponding {\rts} are in Fig. \ref{fig:dqbc}. As shown in Tables \ref{tab:bcr}, \ref{tab:bcl}, \ref{tab:cpmass} and Fig. \ref{fig:dqbc}, our results are in accordance with other theoretical results.

In the $B_c$ meson case, there is mixing of the singlet and triplet states; for example, the P-wave states are linear combinations of $^3P_1$ and $^1P_1$ which can be described by $P'=^1P_1\,cos\theta+^3P_1\,sin\theta$, $P=-^1P_1\,sin\theta+^3P_1\,cos\theta$ \cite{Godfrey:2004ya}. In the diquark $(bc)$ case, there will be the $^3l_l-^1l_l$ mixing via the spin-orbit interaction \cite{Ebert:2002ig} or some other mechanism. We do not consider this kind of mixed state here.

\section{Conclusions}\label{sec:conc}

As shown in Sec. \ref{sec:rtdiquark}, the spectra of the doubly heavy diquarks $(cc)$, $(bb)$, and $(bc)$ obtained by the {\rt} approach agree with other theoretical results. This demonstrates that the {\rt} relation (\ref{massf}), which is appropriate for mesons, baryons and tetraquarks, is suitable for the doubly heavy diquarks.

We present a method to determine the parameters in the diquark {\rts}. Once the parameters are determined, the {\rt} becomes a new and very simple approach for estimating the spectra of diquarks. By employing (\ref{massf}) with (\ref{massform}) and (\ref{cxcons}) to fit the meson {\rts}, we can obtain values of the universal parameters. By fitting a chosen meson {\rt}, $c_{0x}$ is calculated. After all parameters are computed, the diquark {\rt} is definite and the spectra of diquarks can be estimated.

The diquark {\rt} is expected to provide a simple method to investigate easily the $\rho-$mode excitations of baryons, tetraquarks and pentaquarks in the diquark picture.

\acknowledgements
We are very grateful to the anonymous referees for the valuable comments and suggestions.

\appendix*
\section{States of diquarks}\label{app:state}

The diquark {\dqs} can be in two different $SU_c(3)$ configurations, $\bar{3}_c$ and $6_c$. The diquark's color wave function is a superposition of these two different $SU_c(3)$ color representations \cite{Anwar:2018sol,Barabanov:2020jvn}:
\bea
\xi_{color}=a\xi_{\bar{3}_c}+b\xi_{6_c},\;a^2+b^2=1.
\eea
In the diquark picture, the diquark should be in {\cltb} with a quark in $3_c$ to form a colorless baryon. In the tetraquark case, a diquark in {\cltb} or in {\cls} together with an antidiquark in $3_c$ or in $\bar{6}_c$ forms a color singlet.

The total wave function for the diquarks {\dqs} is written as
\bea\label{dwf}
\psi_D=\xi_{color}\otimes\eta_{flavor}\otimes\chi_{spin}\otimes\phi_{space},
\eea
where $\xi_{color}$, $\eta_{flavor}$, $\chi_{spin}$ and $\phi_{space}$ are the color, flavor, spin and spatial wave functions, respectively, see Table \ref{tab:br}. If the quarks are of the same flavor $q=q'$, the diquark wave function $\psi_D$ must be completely antisymmetric to satisfy the Pauli principle.
When a diquark contains light quarks with flavors u,d,s, the overall state $\psi_D$ must also be antisymmetric because strong interactions do not distinguish the flavor of u,d,s \cite{Esposito:2016noz}.

\begin{table}[!phtb]
\caption{Wave functions are symmetric (S) or antisymmetric (A). The flavor wave function for the diquark $(qq')$ can be symmetric $\{qq'\}$ or antisymmetric $[qq']$. $s$ stands for the total spin of two quarks. $n=0,1,2,\cdots$.}  \label{tab:br}
\centering
\begin{tabular*}{0.47\textwidth}{@{\extracolsep{\fill}}lllll@{}}
\hline\hline
    & $\xi_{color}$     &  $\eta_{flavor}$  & $\chi_{spin}$  & $\phi_{space}$   \\
\hline
S    & $6_{c}$           & $\{qq'\}$         & $s=1$          & $l=2n$    \\
A    & $\bar{3}_c$       & $[qq']$           & $s=0$          & $l=2n+1$   \\
\hline\hline
\end{tabular*}
\end{table}

In the color space, the color wave function can be analyzed by employing the $SU_c(3)$ group theory
\bea
3_c\otimes3_c={\bar{3}_c}\oplus6_c.
\eea
The color wave functions for the diquark read as
\bea
\xi_{\bar{3}_c}=|(qq')^{\bar{3}_c}\rangle,\quad
\xi_{6_c}=|(qq')^{6_c}\rangle.
\eea
The spin triplets are symmetric while the spin singlet is antisymmetric. The spin wave functions read as
\bea
\chi_{S}=|(qq')_1\rangle,\quad \chi_{A}=|(qq')_0\rangle.
\eea
If $q{\ne}q'$, the flavor wave functions can be written in the symmetric and the antisymmetric form as
\begin{align}
\eta_S&=\{qq'\}=\frac{1}{\sqrt{2}}(qq'+q'q),\nonumber\\
\eta_A&=[qq']=\frac{1}{\sqrt{2}}(qq'-q'q)
\end{align}
while
\bea
\eta_{flavor}=\eta_S=\{qq'\}=qq
\eea
if $q=q'$.

Since quarks have the same intrinsic parities, the overall parity is
\bea
P((qq'))=(-1)^l,
\eea
where $l$ is the symmetry of the orbital wave function $\phi_{space}$.

\begin{table*}[!phtb]
\caption{The completely antisymmetric states for the diquarks in {\cltb} and in $6_c$. $j$ is the spin of the diquark {\dqs}, $s$ denotes the total spin of two quarks, $l$ represents the orbital angular momentum. $n=n_r+1$, $n_r$ is the radial quantum number, $n_r=0,1,2,\cdots$.}  \label{tab:dqstates}
\centering
\begin{tabular*}{0.8\textwidth}{@{\extracolsep{\fill}}ccccc@{}}
\hline\hline
 Spin of diquark & Parity  &  Wave state  &  Configuration    \\
( $j$ )          & $(j^P)$ & $(n^{2s+1}l_j)$  \\
\hline
j=0              & $0^+$   & $n^1s_0$         & $[qq']^{{\cltba}}_{n^1s_0}$,\; $\{qq'\}^{{6_c}}_{n^1s_0}$ \\
                 & $0^-$   & $n^3p_0$         & $[qq']^{{\cltba}}_{n^3p_0}$,\; $\{qq'\}^{{6_c}}_{n^3p_0}$       \\
j=1              & $1^+$   & $n^3s_1$, $n^3d_1$   & $\{qq'\}^{{\cltba}}_{n^3s_1}$,\;    $\{qq'\}^{{\cltba}}_{n^3d_1}$,\;
$[qq']^{6_c}_{n^3s_1}$,\;    $[qq']^{{6_c}}_{n^3d_1}$\\
                 & $1^-$   & $n^1p_1$, $n^3p_1$   &
$\{qq'\}^{{\cltba}}_{n^1p_1}$,\; $[qq']^{{\cltba}}_{n^3p_1}$, \;          $[qq']^{6_c}_{n^1p_1}$,\; $\{qq'\}^{6_c}_{n^3p_1}$ \\
j=2              & $2^+$   & $n^1d_2$, $n^3d_2$         &  $[qq']^{{\cltba}}_{n^1d_2}$,\; $\{qq'\}^{{\cltba}}_{n^3d_2}$,\;
$\{qq'\}^{6_c}_{n^1d_2}$,\; $[qq']^{6_c}_{n^3d_2}$         \\
                 & $2^-$   & $n^3p_2$, $n^3f_2$       &
 $[qq']^{{\cltba}}_{n^3p_2}$,\; $[qq']^{{\cltba}}_{n^3f_2}$,\;
  $\{qq'\}^{6_c}_{n^3p_2}$,\; $\{qq'\}^{6_c}_{n^3f_2}$          \\
$\cdots$         & $\cdots$ & $\cdots$               & $\cdots$  \\
\hline\hline
\end{tabular*}
\end{table*}

For the diquarks composed of two identical quarks, the antisymmetric flavor wave function does not exist; therefore, the states in the $[qq]$ configuration disappear in Table \ref{tab:dqstates}. From Table \ref{tab:dqstates}, we can easily read the possible mixing of different states. For example, for the diquarks composed of different quarks, the $1^+$ state can be a mixture of the S-wave state and D-wave state, and the mixture of the {\cltb} state and $6_c$ state.


\begin{thebibliography}{99}
\bibitem{note}
The {\rts} are commenly plotted in the $(M^2,\,x)$ plane or in the $(x,\,M^2)$ plane, where $x=l,\,n_r$. For simplicity, the figures plotted in the $(M,\,x)$ plane, in the $(M-m,\,x)$ plane and in the $((M-m)^2,\,x)$ plane are also called the {\rts}.
\end{thebibliography}
\end{document}